\documentclass[twocolumn,tighten,times,astrosymb,floatfix]{aastex631}

\usepackage{
	amsmath,
	amssymb,
	amsthm,
	booktabs,
        breqn,
	epsfig,
	graphicx,
	grffile,
	import,
	morefloats,
	microtype,
	natbib,
	subfigure,
	ulem,
	url,
	verbatim,
	wasysym,
	xcolor,
	xspace,
}
\usepackage{savesym}
\savesymbol{tablenum}
\usepackage{siunitx}
\restoresymbol{SIX}{tablenum}
\usepackage[version=4]{mhchem}
\usepackage[mathlines]{lineno}
\usepackage{lineno}

\DeclareSIUnit{\jy}{Jy}
\DeclareSIUnit{\beam}{beam}
\DeclareSIUnit{\kms}{\kilo\meter\per\second}

\shorttitle{Terrestrial zone gas around HD 131488}
\shortauthors{Worthen et al.}

\begin{document}

\title{\Large
Kinematics of terrestrial zone CO gas in the HD 131488 debris disk from CRIRES+
}

\correspondingauthor{Kadin Worthen}
\email{kworthe1@jhu.edu}

\author[0000-0002-5885-5779]{Kadin Worthen}
\affiliation{William H. Miller III Department of Physics and Astronomy, John's Hopkins University, 3400 N. Charles Street, Baltimore, MD 21218, USA}
\affiliation{NASA Postdoctoral Program Fellow, NASA Goddard
Space Flight Center, 8800 Greenbelt Rd, Greenbelt, MD 20771, USA}

\author[0000-0002-8382-0447]{Christine H. Chen}
\affiliation{Space Telescope Science Institute, 3700 San Martin Drive, Baltimore, MD 21218, USA }
\affiliation{William H. Miller III Department of Physics and Astronomy, John's Hopkins University, 3400 N. Charles Street, Baltimore, MD 21218, USA}

\author[0000-0001-9352-0248]{Cicero X. Lu}
\affiliation{Gemini Observatory/NSF’s NOIRLab, 670N. A’ohoku Place, Hilo, HI 96720, USA}

\author[0000-0002-9274-8715]{Kevin D. Smith}
\affiliation{School of Physics, Trinity College Dublin, College Green, Dublin 2, Ireland}

 \author[0000-0002-5758-150X]{Joan R. Najita}
 \affiliation{ NSF’s NOIRLab, 950 N. Cherry Avenue, Tucson, AZ 85719, USA}

 \author[0000-0003-4705-3188]{Luca Matr\`a}
 \affiliation{School of Physics, Trinity College Dublin, the University of Dublin, College Green, Dublin 2, Ireland}

\author[0000-0001-5638-1330]{Sean D. Brittain}
\affiliation{Department of Physics and Astronomy, 118 Kinard Laboratory, Clemson University, Clemson, SC 29634, USA}

\author[0000-0002-4388-6417]{Isabel Rebollido}
\affiliation{European Space Agency (ESA), European Space Astronomy Centre (ESAC), Camino Bajo del Castillo s/n, 28692 Villanueva de la Cañada, Madrid, Spain}

\author[0000-0002-4803-6200]{A. Meredith Hughes}
\affiliation{Department of Astronomy, Van Vleck Observatory, Wesleyan University, 96 Foss Hill Drive, Middletown, CT 06459, USA}

\author[0009-0007-3677-8040]{Chia-Ying Chung}
\affiliation{William H. Miller III Department of Physics and Astronomy, John's Hopkins University, 3400 N. Charles Street, Baltimore, MD 21218, USA}

\author[0000-0002-2989-3725]{Aki Roberge}
\affiliation{NASA Goddard Space Flight Center, Greenbelt, USA}

\begin{abstract}
We present CRIRES+ M4368 observations of HD 131488, which hosts one of the most CO-rich debris disks. With these observations, we detect and spectrally resolve more than 10 fluorescently excited fundamental ro-vibrational CO emission lines from the HD 131488 circumstellar disk. We find that CO is warm, with a rotational temperature of $\sim$200-300 K. We extract a spectro-astrometric signal from the CO emission and model it together with the CO line profiles to determine the inner radial extent of the warm CO. This kinematic modeling suggests an inner radius for the emitting fluoresced CO of 3.5$\pm$0.3 au. We explore potential origins for this warm CO and model the radial evolution of the CO in this system under the assumption that the gas is of secondary origin. We find, with caveats, that the inner gas probed by CRIRES+ could be the result of viscous spreading of CO released in the outer planetesimal belt as long as the gas viscosity corresponds to $\alpha>10^{-4}$. These radial evolution models predict that the CO will spread inward to the central star, but our CRIRES+ observations do not show evidence for gas interior to 3 au. We discuss potential explanations for this discrepancy such as unseen planets accreting the gas and creating a cavity in the disk. We alternatively discuss a primordial origin for the gas in this system where the inner CO depleted region ($<$3.5 au) may be due to the dispersal of the primordial gas. 

\end{abstract}



\section{Introduction}\label{sec:Introduction}
Debris disks are dust dominated circumstellar disks whose dust is continuously replenished by collisional processes involving the destruction of planetesimals \citep{Hughes18}. Such second-generation debris disks are thought to exist after the dispersal of the gas dominated protoplanetary disk \citep{Wyatt15}. In this traditional picture of disk evolution, debris disks were thought to be depleted in gas.  Millimeter observations of debris disks, however,  have detected carbon monoxide in $\sim$20 debris disks (e.g. \citealt{Moor11,Sifry16,Macgregor18,Rebollido22,Moor25}).  The range of CO masses inferred from millimeter observations of CO in debris disks spans multiple orders of magnitude from $\sim10^{-7}$-$10^{-1}$ M$_{\oplus}$ \citep{Matra17b,Moor17}.

For the lowest CO mass debris disks, it is thought that the gas must be constantly replenished and is thus of secondary origin, because, even in a primordial origin scenario with H$_2$/CO=10$^4$, there would not be a high enough column density of gas to shield CO against photodissociation to the current ages of the host stars \citep{Marino16, Matra17, Matra17b}. For the debris disks with the highest CO masses ($\gtrsim10^{-2}$ M$_{\oplus}$), the origin of the gas is not clear as it could be of secondary origin like the dust (e.g. \citealt{Marino20}) or it could be long lived, H$_2$ dominated, primordial gas from the parent protoplanetary disk \citep{Nakatani23}. Some systems have been proposed to be hybrid disks, where they possess second-generation dust while still harboring primordial gas \citep{Kospal13,Worthen26}. Answering the question on the origin of the gas in debris disks is of high interest as such long lived H$_2$-rich primordial gas could increase the timescale of gas giant planet formation out to the 15-40 Myr old ages (e.g. \citealt{Zuckerman12}) of these CO rich disks.

Large masses of second-generation gas would also have important implications for the formation and evolution of planetary systems. Simulations of second-generation gas production and evolution have suggested that secondary gas can viscously spread into the terrestrial planet zone of planetary systems and accrete onto young planets \citep{Kral20}. These models show that an Earth's atmosphere mass of gas could be accreted onto rocky planets for low CO production rates (10$^{-8}$ M$_{\oplus}$/Myr) and secondary atmospheres with pressures comparable to sub-Neptunes can be accreted for high CO production rates (10$^{-2}$ M$_{\oplus}$/Myr) \citep{Kral20}. This accretion of second-generation gas may not only form secondary atmospheres on rocky planets, but can also change the composition of primordial atmospheres \citep{Kral20}. For example, the high carbon abundances in the atmospheres of Uranus and Neptune have been proposed to be due to such secondary gas accretion from circumstellar gas produced by a young Kuiper Belt \citep{Huet25}. 

If second-generation gas is seeding terrestrial planets with secondary atmospheres, as predicted by \cite{Kral20}, then we would expect to find gas in the terrestrial planet zone ($\sim$ inner few au) of debris disk systems. Most of the CO detected in debris disks at millimeter wavelengths by ALMA, however, exists at stellocentric distances greater than 10 au (\citealt{Cataldi23} and references therein). There are two debris disk systems, HD 172555 and HD 23514, that exhibit molecular gas in their terrestrial planet regions where the gas is thought to be due to large impacts \citep{Schneiderman21, Su25}. In younger, gas-rich protoplanetary disks that harbor primordial gas, fundamental ($\Delta v=1$) ro-vibrational CO emission at 4.7 $\mu$m that arises from the inner few au of those disks is routinely detected \citep{Najita03,Brittain07,Brown13,Dickson-Vandervelde25}. In debris disks, fundamental ro-vibrational CO emission has been detected in three sources. The first is 49 Ceti, where the ro-vibrational CO lines are fluorescently excited from stellar photons and are spatially resolved at stellocentric distances between $\sim50$ to 120 au with JWST \citep{Worthen25}. Those JWST observations of 49 Ceti do not cover the central star, so it is not known how far in the ro-vibrational CO emission extends. The second is $\beta$ Pic, where the CO emission is also spatially resolved by JWST beyond 50 au \citep{Betti26}. The other debris disk that is known to have fluorescently excited ro-vibrational CO emission is the one around HD 131488. For this system, JWST detected fundamental CO emission lines in the spectrum of the unresolved point source and the CO excitation temperature suggests that it may arise from the inner few au \citep{Lu26}. In this work, we present CRIRES+ observations to kinematically determine the location of the warm CO detected in \cite{Lu26}. 

HD 131488 is an A-type star that is a member of Upper Centaurus Lupus (UCL) within the Scorpius-Centaurus OB association and it has an age of $\sim$16 Myr \citep{Preibisch08, Pecaut12, Pecaut16}. It is host to one of the most CO-rich debris disks with a CO mass, based on ALMA observations of C$^{18}$O and assuming an interstellar medium (ISM) isotopologue ratio, of 8.9$\times10^{-2}$ M$_{\oplus}$ \citep{Moor17}. High-resolution ALMA observations revealed that the CO has a broad spatial extent from a stellocentric distance of 6 au out to 160 au \citep{MacManamon26}. Optical spectra of HD 131488 do not detect any emission lines or filling of Balmer absorption lines in the stellar spectra that could be associated with accretion of hydrogen gas onto the star \citep{Melis13}. However, low levels of accretion may still remain undetected against the stellar photosphere. The cold CO gas component of the HD 131488 disk has also been detected with absorption line spectroscopy in rovibronic transitions of CO at UV wavelengths with HST \citep{Brennan24} as well as first-overtone ($\Delta v=2$) transitions of CO in K-band with CRIRES+ \citep{Smith25}. The CRIRES+ K-band observations searched for H$_2$ absorption along the line of sight with the aim to determine if the gas composition indicates a primordial or secondary origin for the gas. Those observations resulted in a non-detection of H$_2$ along the line of sight but placed a limit on the CO/H$_2$ ratio of CO/H$_2 >3.09\times10^{-5}$, which does not completely rule out a primordial gas composition \citep{Smith25}. The origin of the gas in this system is then still ambiguous. Properties of the HD 131488 system are shown in Table \ref{tab:system_params}.

Along with the CO-rich disk around HD 131488, SED modeling that includes near- and mid-IR spectroscopy and photometry indicates that the debris disks of HD 131488 contains both a population of cold (100 K) and warm (750 K) dust \citep{Melis13}. It was proposed that the hot dust could be due to collisions involved in the final growth phase of terrestrial planets \citep{Melis13}. If there are rocky planet embryos in the inner regions of this system, then it is of interest to understand if the CO gas disk extends inwards to the terrestrial planet zone as it could be a potential source of gas that can be accreted by young rocky planets as proposed by \cite{Kral20}. 

The existing JWST/NIRSpec observations of ro-vibrational CO in HD 131488 had a spectral resolution of R$\sim$3,000 so the CO emission lines were spectrally unresolved and kinematic information on the gas location was limited \citep{Lu26}. The model that reproduced the observed CO emission spectrum from JWST had the CO extend inwards to sub-au distances \citep{Lu26}, but without spectrally resolved  kinematics, the radial distribution of this fluorescently excited warm CO could not be confirmed. To that end, we present CRIRES+ observations of the ro-vibrational CO emission lines from the HD 131488 disk. In Section 2, we describe our observations and data reduction. In Section 3 we describe our CO line detections and kinematic modeling. In Section 4 we discuss possible explanations for the inner warm CO and in Section 5 we present our conclusions.

\begin{table}[htbp!]
    \centering
    \caption{Stellar and Disk Parameters for HD 131488}
    \begin{tabular}{c c c}
    \hline
     Parameter & HD 131488 & Reference  \\
     \hline
     Spectral Type &A1V & (1)  \\
     T$_{\mathrm{eff}}$& 8700$^{+160}_{-210}$ K &(2)  \\
     L$_{*}$& 11.8 L$_{\odot}$ & (3)   \\
     M$_{*}$ (Photometric)& 1.804$\pm0.016$ M$_{\odot}$ & (4)   \\
     Distance & 152 pc & (5)\\
     System Age & $\sim$16 Myr & (6)  \\
     Disk CO Mass & 8.9$\times10^{-2}$ M$_{\oplus}$ & (7)   \\
     CO R$_{in}$& 3.5 $^{+0.3}_{-0.3}$ au & (8)  \\
     Disk Inclination & 84.87$^{+0.05}_{-0.05}$$^{\circ}$& (9) \\
     Radial Velocity & 5.97 $\pm0.05$ km/s &(10) \\
    \hline
    \end{tabular}
         
                \begin{minipage}{6 cm}
    \vspace{0.2cm}
    
    \textbf{References.} (1). \citealt{Houk78} (2). \citealt{Brennan24} (3). \citealt{Matra25} 
    (4). \citealt{Marino26} (5). \citealt{Pecaut16} (6). \citealt{Gaia23} (7). \citealt{Moor17} (8). This work (9). \citealt{Zawadski26} (10). \citealt{Smith25}
    \end{minipage}

    \label{tab:system_params}
\end{table}

\section{Observations and Data Reduction}\label{sec:Observations}
\subsection{Description of Observations }
We observed HD 131488 (K-band magnitude of 7.803) using the 0.2'' slit (R$\sim$ 100,000) on CRIRES+ \citep{Dorn23} on the Very Large Telescope (VLT) in 4 $\sim$1-hour observing blocks across 3 different nights: July 15, 20, and 24 2025 (Program 115.2851.001, PI K. Worthen). We observed HD 131488 in the M4368 setting to cover the wavelength range between 4.683 and 4.777 $\mu$m, which includes the brightest fundamental ro-vibrational CO emission lines detected by JWST/NIRSpec \citep{Lu26}. Our observing dates were chosen such that the CO lines of interest would be shifted in velocity, due to the orbital motion of Earth, away from nearby telluric lines by a minimum velocity of 15 km s$^{-1}$. The observations were taken with the adaptive optics system for CRIRES+ employed. We oriented the CRIRES+ slit such that it was aligned with the position angle of the disk (96$^{\circ}$ east of north) determined from modeling of ALMA observations of the millimeter dust continuum from \cite{Worthen24A}. 

Because of the high number of telluric lines in this M-band CRIRES+ setting, we immediately followed the observations of HD 131488 in each observing block by observations of $\beta$ Lup (a B2IV star with a K-band magnitude of 3.32), which we used as a telluric standard to remove the telluric features from the observations of HD 131488.

The CRIRES+ observations were taken in an ABBA nodding pattern with a nod throw of 6 arcseconds along the slit. An overview of the observational parameters is shown in Table \ref{tab:obs_params}. The relatively small differences in airmass between the telluric standard and HD 131488, makes the $\beta$ Lup observations a good representation of the telluric lines in our science data as airmass differences less than 0.2 between standard star and science star are considered suitable \cite{Ulmer-Moll19}. 

\begin{table*}[htbp!]
    \centering
    \caption{Overview of CRIRES+ Observations}
    \begin{tabular}{c c c c c c c }
    \hline
     Night & Target & Exposure Time (s) & Airmass  & Wavelength Setting & PWV (mm) & Slit P.A. ({$^\circ$}) \\
     \hline
     2025 July 15&HD 131488 &2160 &1.063 &M4368 &1.78 &96  \\
     2025 July 15&$\beta$ Lup &60 &1.130 &M4368 &2.11 &96  \\
     2025 July 20&HD 131488 &2160 &1.044 &M4368&0.80 & 96  \\
     2025 July 20&$\beta$ Lup &60 &1.083 &M4368 &0.79 &96 \\
     2025 July 24&HD 131488 &2160 &1.062 &M4368 &1.27 & 96  \\
     2025 July 24&$\beta$ Lup &60 &1.130 &M4368 &1.35 & 96  \\
     2025 July 24&HD 131488 &2160 &1.143 &M4368 &1.35 &  96 \\
     2025 July 24&$\beta$ Lup &60 &1.267 &M4368 &1.31 &96   \\
    \hline
    \end{tabular}
         
                \begin{minipage}{13 cm}
    \vspace{0.2cm}
    
     \textbf{Note.} PWV is the precipitable water vapor in mm at the start of the observations. We had two observing blocks on the same night of July 24.
    \end{minipage}

    \label{tab:obs_params}
\end{table*}

\subsection{Data Reduction}
We used ESO CR2RES EsoReflex pipeline (version 1.6.7) \citep{Freudling13} with default reduction parameters to reduce the 2D detector images and extract separate spectra for the A and B nods from all three CRIRES+ chips for both HD 131488 and $\beta$ Lup. We did not use the combined A and B nod extracted spectra provided by the pipeline because it is well known that the the A and B nod spectra can have wavelength offsets when the star does illuminate the full width of the slit, which was the case with our observations  (e.g. \citealt{Grant24, Smith25}). We therefore used the separate  A and B nod spectra extracted by the EsoReflex pipeline, and we corrected for this offset for each night and each  chip before combining the A and B nods from each night. 

\begin{figure}[htbp!]
    \centering
    \includegraphics[scale=0.5]{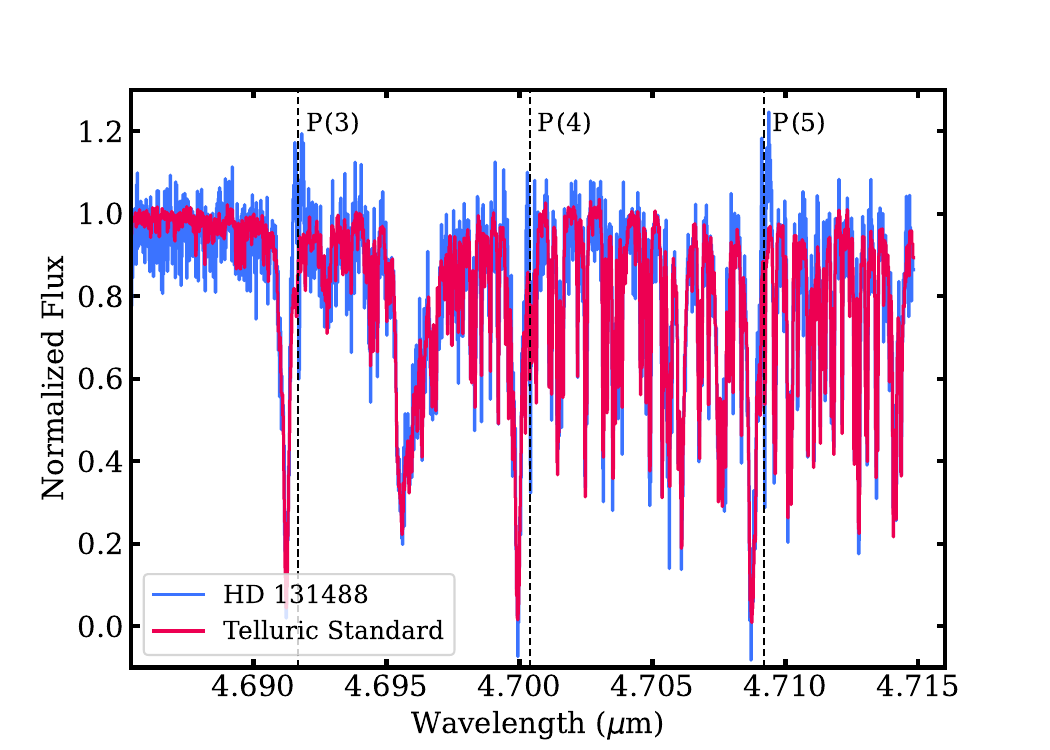}
    \caption{A-nod CRIRES+ spectrum from chip 1 of both HD 131488 (blue) and the telluric standard (red) from our first observing block. The majority of absorption features shared between both spectra are telluric lines. The vertical black dashed lines show the expected wavelength of CO ro-vibrational emission (v=1-0) lines after correcting for the orbital velocity of Earth relative to HD 131488 on the night of the observations and the radial velocity of the star \citep{Smith25}. The P(3) and P(5) CO emission lines are visibly present in HD 131488 while absent in the telluric standard. }
    \label{fig:Tell_spec}
\end{figure}

We did this wavelength correction by cross correlating the non-telluric-corrected A and B nod spectra and finding the wavelength shift that corresponds to the maximum correlation value between the two nod positions. This offset was determined separately for HD 131488 and the telluric standard for each of the 4 observing blocks and 3 chips. The shift in wavelength between the A and B nods for both the telluric standard and HD 131488 across all observing blocks ranged between 0.06 and 0.075 nm. One A-nod spectra for HD 131488 and the telluric standard, from the first observing block, are shown in Figure \ref{fig:Tell_spec}. This figure demonstrates that the v=1-0 P(3) and P(5) CO emission lines are identifiable in the HD 131488 spectrum prior to any telluric correction. 

We did not find a wavelength offset when comparing the A and B nod positions of HD 131488 to the corresponding nod positions of the telluric standard from the same observing block (cross correlation value peaked at zero offset). To correct the tellurics from the HD 131488 spectra, we  divided the telluric standard spectrum from HD 131488's spectrum for each nod position, shifted the B nod spectra by the wavelength offset determined from cross correlation, and then median combined the A and B-nod spectra. This was done separately for each observing block and CRIRES+ chip. This gave telluric-corrected combined A+B-nod spectra for each observing block and chip. Because the four observing blocks were taken over the span of nine days, the orbital velocity of Earth relative to HD 131488 changed between observing blocks, making it necessary to shift and align the spectra from each block before combining the data into one final spectrum. We did this by fitting Gaussians to the astrophysical CO absorption lines in the telluric-corrected spectra from each observing block, computing the wavelength offset between the line centers in each observing block, and then shifting all spectra to the velocity of the first observing block (the night of July 15, 2025). Once all spectra from the observing blocks were aligned, we median combined all four spectra for each chip to produce a final telluric-corrected and aligned non-continuous spectrum from $\sim$4.68 to 4.78 $\mu$m. To flux calibrate the CRIRES+ spectrum, we pinned it to the continuum flux of the JWST/NIRSpec spectrum \citep{Lu26} in a CO line-free region at 4.6865 $\mu$m. The final telluric-corrected continuum-subtracted spectrum of HD 131488 is shown in Figure \ref{fig:Final_spec}. We estimated the continuum to be subtracted by measuring the median continuum in  the line free regions for each chip.

\subsection{Spectro-astrometry}
We also searched for a spectro-astrometric (SA) signal in our data. A SA signal is a shift in the centroid of the PSF as a function of wavelength (e.g. \citealt{Bailey98}) due to, in this case, the circumstellar CO emission in the HD 131488 disk. In the case of gas in a circumstellar disk, the red- and blue-shifted sides of the emission line will have centroid offsets that extend in opposite directions in the spatial direction of the slit axis when the position angle of the slit is aligned with the position angle of the disk \citep{Pontoppidan08,Jensen24}. A SA measurement can then be used to check if the gas is in Keplerian orbit around the star  as well as search for asymmetric structures in the inner regions of gas disks \citep{Brittain15, Jensen24}. Typically, the centroid of the PSF can be measured to spatial offsets that are a fraction of the telescope's angular resolution, enabling the measure of the center of light at a given wavelength at scales smaller than the angular resolution from the PSF \citep{Pontoppidan11,Brittain15}. For our observations of HD 131488, we only observed one slit position angle, but the slit is aligned with the position angle of the disk, giving us the maximum SA signal. 

\begin{figure*}
    \centering
    \includegraphics[scale=0.75]{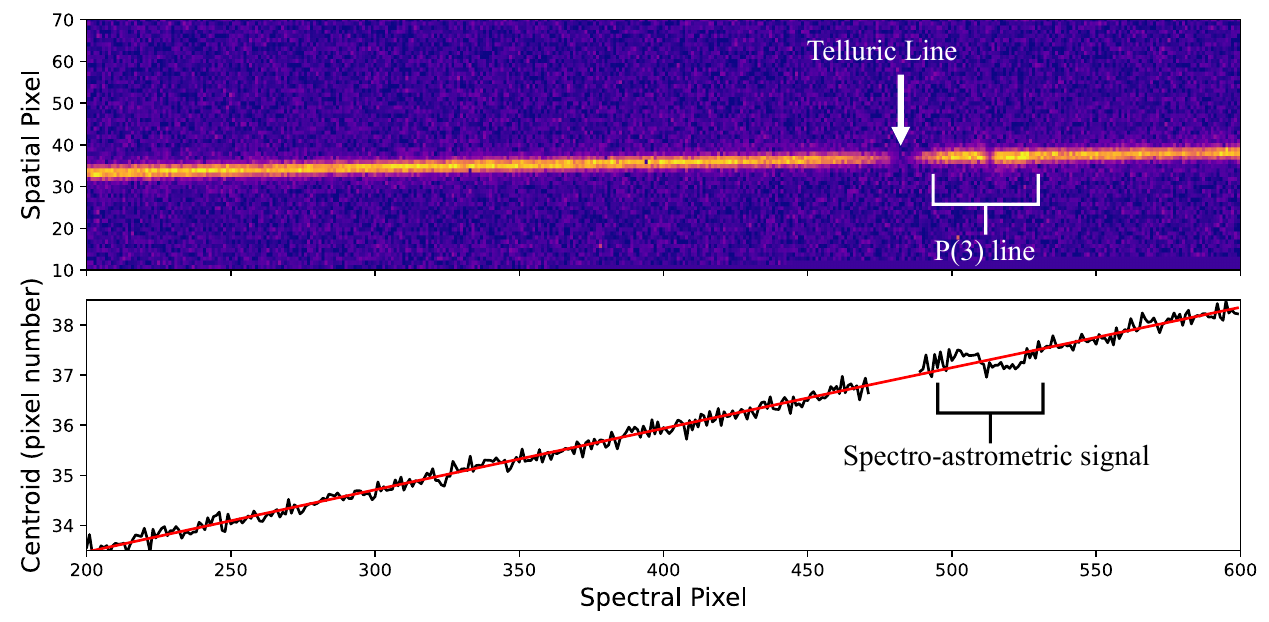}
    \caption{Top: CRIRES+ 2D detector image of the A-nod from observing block 2 showing the spectral trace and the location of the $^{12}$CO v=1-0 P(3) line as well as the nearby telluric line. Bottom: Centroid of the PSF from the Gaussian fitting as a function of pixel number in the spectral direction (wavelength). The red line shows the best fit spline to regions outside of the SA signal that is used to subtract off this underlying continuum of the trace. The region that contains the SA signal that was not used in the spline fit is the range of data from spectral pixels 494 to 534 as shown by the SA signal label. The centroid extensions above and below the red continuum at the location of the P(3) line is the SA signal. }  
    \label{fig:spec_ast}
\end{figure*}

To extract the SA signal, we first located the position of the CO emission lines in the 2D detector images. We then fit a Gaussian to the PSF in each column in the dispersion direction to measure the centroid of the PSF as a function of wavelength. The 2D detector image from observing block 2 with the location of the $^{12}$CO v=1-0 P(3) line labeled as well as the centroid measurement as a function of spectral pixel (wavelength) from  this detector image are shown in Figure \ref{fig:spec_ast}. Once we measured the PSF centroid across the wavelength range of interest, we fit a cubic spline (3rd order with 2 knots) to the regions outside of the SA signal, interpolated the spline to the wavelengths that contain the SA signal, and then subtracted off the spline that represents the underlying spectral trace to leave just the SA signal. An example of the best fit spline to the spectral trace is also shown in Figure \ref{fig:spec_ast}. We repeated this same process for the A and B nods of each observing block and for all four observing blocks. We then shifted and aligned the extracted SA signal from each observing block and A/B nods based on the wavelength shifts determined between nods and observing blocks for the 1D spectra above. The final SA signal is shown below in section 3.2. We only include the P(3) CO line in the SA signal because it is the line with the highest S/N ratio for its SA signal and we find that including other lines with lower S/N decreases the overall S/N ratio of the SA profile. 

To check the robustness of the SA signal, we performed the same centroid measurements to search for a false SA signal on the calibrator star $\beta$ Lup, as recommended by \citealt{Brittain15}. We found that the PSF centroid as a function of wavelength for the calibrator star shows no SA signal greater than 3$\sigma$ at the same wavelengths that we detected the SA signal in the HD 131488 data. Another check for the robustness of a SA signal is to measure the photocenter of light in telluric lines where no SA is expected \citep{Davies10,Whelan15}. We searched for false SA signals in telluric features between 4.7 and 4.705 $\mu$m in our HD 131488 data and similarly found no false SA signals at the 3$\sigma$ level. It is also possible that the slit occults some of the emission from the source if the slit is not well centered, which can induce an asymmetry onto the observed SA signal \citep{Brittain15}. We, however, found that our SA signal is symmetric and repeatable over all 4 observing blocks. These checks support that the SA signal we measured in the HD 131488 CRIRES+ observations is likely not an artifact.

\section{Results}
\subsection{Line Detections}
\begin{figure*}[htbp!]
    \centering
    \includegraphics[scale=0.57]{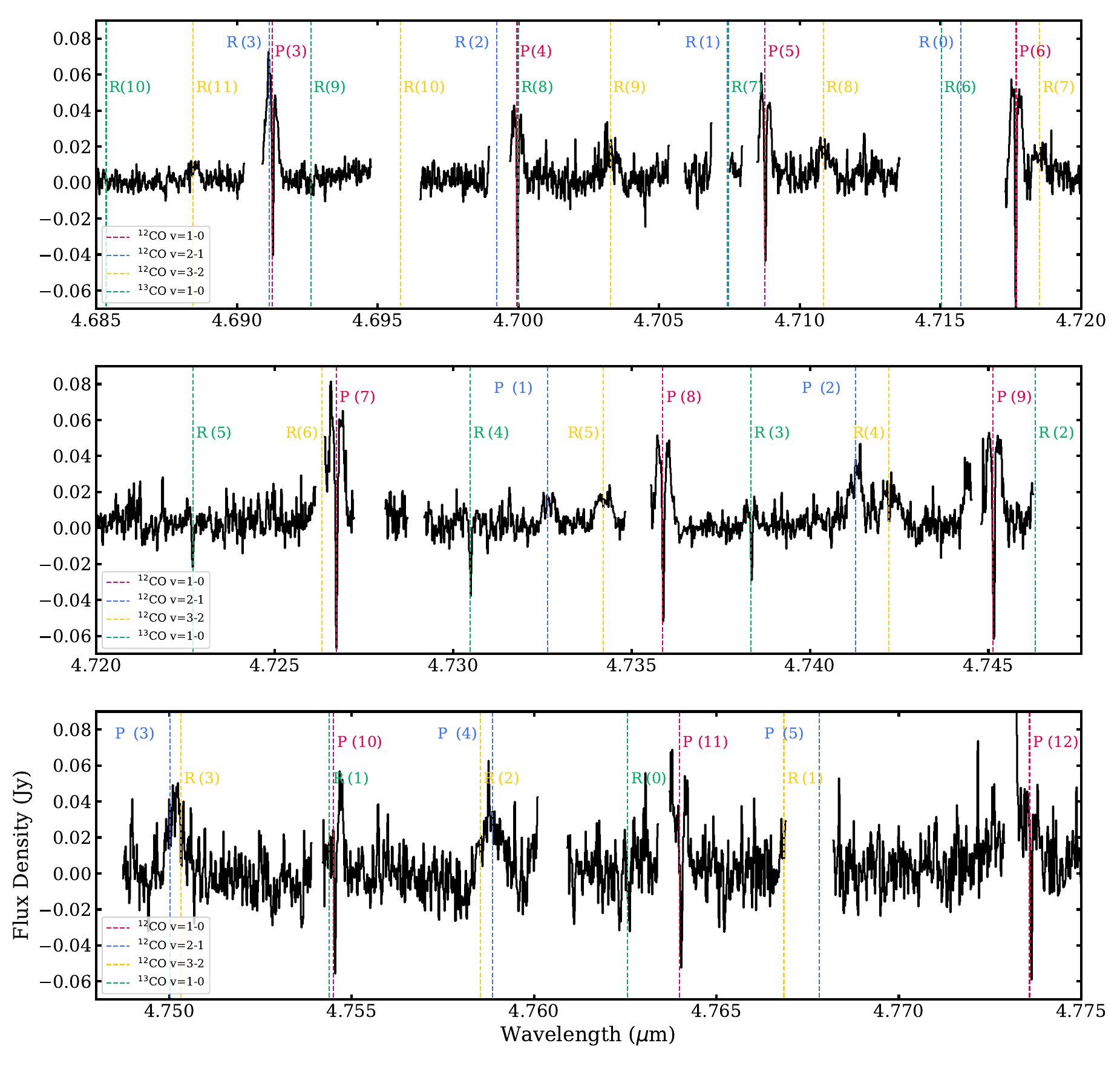}
    \caption{Telluric-corrected and continuum subtracted CRIRES+ spectrum of HD 131488 after combining all four observing blocks of data. The regions of high telluric contamination are masked out. The vertical dashed lines show the location of $^{12}$CO v=1-0  (red), $^{12}$CO v=2-1 (blue), $^{12}$CO v=3-2 (yellow), and $^{13}$CO v=1-0 (green) transitions. The spectrum shown is shifted in wavelength by correcting for the orbital velocity of Earth on the night of the observations ($\sim$22 km s$^{-1}$) and the radial velocity of the star \citep{Smith25}. The labels for each transition indicate whether it is an R- or P-branch line.}
    \label{fig:Final_spec}
\end{figure*}

We detect the v=1-0 P(3), P(4), P(5), P(6), P(7), P(8), and P(9) lines of $^{12}$CO in emission from the HD 131488 disk as shown in Figure \ref{fig:Final_spec}. Each of those lines also has an absorption component super imposed on the emission line, at least in part, due to absorption from cold gas in the outer disk along the line of sight to the star. There could also be a contribution of self-absorption from the emitting region on top of the absorption from the outer disk. Such absorption from the outer disk against the stellar photosphere was previously detected in K-band observations of first overtone ($\Delta v=2$) of CO \citep{Smith25}. Our CO absorption lines similarly absorb the stellar photosphere, so this cold gas along the line of sight contributes in part to the absorption detected here. The K-band observations, however, only detect rotational transitions as large as J=8-7. We detect up to the P(12) line in absorption, potentially hinting at different temperatures and thus some contribution from self absorption. A model of the excitation of the absorption lines seen in this M-band data will be necessary to determine if self-absorption provides a contribution to the absorption lines. This modeling is left for future work. The P(10), P(11), and P(12) lines are all detected in absorption, however, their emission lines are strongly affected by tellurics and none of these three lines have complete emission line profiles as a result (see the bottom panel of Figure \ref{fig:Final_spec}). Because of this we only use the P(3)-P(9) lines of the v=1-0 transitions for the remainder of the analysis. 

Along with the $^{12}$CO v=1-0  lines, we also detect multiple $^{12}$CO v=2-1 and v=3-2 lines in emission and $^{13}$CO v=1-0 lines in absorption. The P(1), P(2), P(3), and P(4) lines of the v=2-1 vibrational transition are detected in emission. The $^{12}$CO v=2-1 R(3)  line overlaps with the $^{12}$CO  v=1-0 P(3)  line, as shown in  Figure \ref{fig:Final_spec}, making those two lines subject to contamination due to blending. We also detect the $^{12}$CO v=3-2 R(3), R(4), R(5), R(7), R(8), and R(9)  lines in emission along with the $^{13}$CO v=1-0 R(3), R(4), and R(5) lines in absorption. We leave the analysis of the $^{13}$CO absorption lines and the CO isotoplogue ratio for future work and here we focus on the CO kinematics. 

The detection of CO emission lines with upper vibrational states $>$1, as seen here, is expected for CO that is fluorescently excited by stellar UV photons when the thermal temperature of that gas is not sufficient to populate these higher vibrational states (e.g. \citealt{Brittain07, Worthen25}).  The modeling of the JWST/NIRSpec spectrum of the CO around HD 131488 showed that UV fluorescence is the likely excitation mechanism of these rovibrational CO emission lines because the rotational temperature is much lower than the vibrational temperature of the CO \citep{Lu26}. This modeling of the JWST spectrum also suggested that the CO emission lines were affected by absorption from cold CO along the line of sight from the edge-on disk. This absorption component had been previously detected at UV and IR wavelengths for this source \citep{Brennan24, Smith25}. The high spectral resolution of CRIRES+ does show a narrow absorption component super imposed on the $^{12}$CO v=1-0 lines, likely due to absorption from colder gas in the outer edge-on disk \citep[$i \sim 85^\circ$,][]{Zawadski26} along the line of sight.

\subsection{Kinematic Modeling}

Each of the$^{12}$CO v=1-0 P(3)-P(9) lines are spectrally resolved with our CRIRES+ observations (R$\sim$3 km s$^{-1}$ ), enabling a kinematic analysis of the gas to determine the location of the CO in the disk, as has been done from high-spectral resolution near-IR observations of CO emission from protoplanetary disks \citep{Pontoppidan08}. The model for HD 131488 from \cite{Lu26} that was used to explain the CO emission around HD 131488 detected with JWST/NIRSpec had the CO go inwards to a stellocentric distance of 0.18 au with the majority of the gas emission coming from a region with a stellocentric distance between 0.6 and 3 au. At the spectral resolution of the JWST observations (R$\sim$3,000), the CO lines were spectrally unresolved, making it not feasible to kinematically determine where this warm CO is in this disk. 

\begin{figure*}
    \centering
    \includegraphics[scale=0.6]{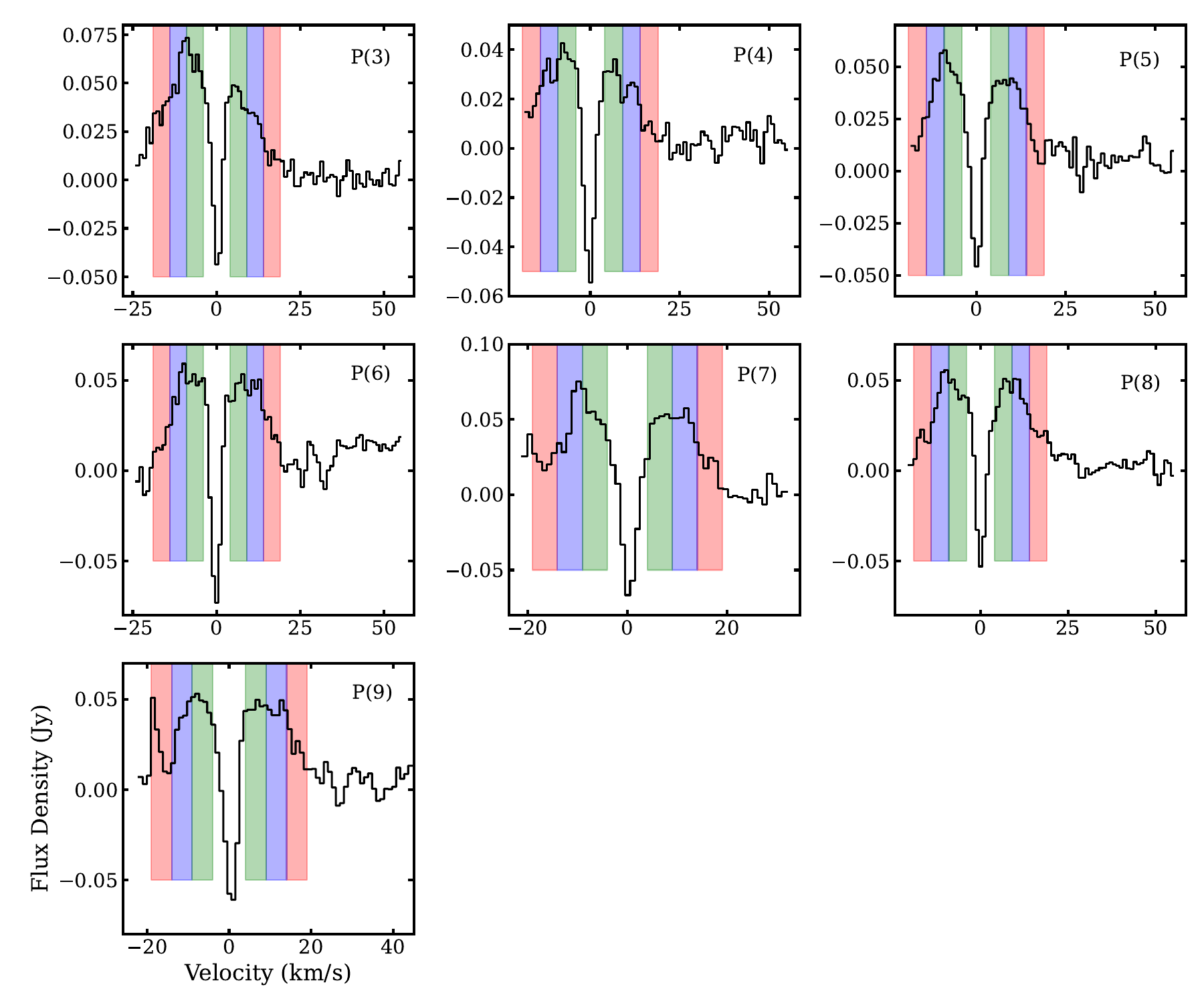}
    \caption{Spectrally resolved continuum-subtracted line profiles for the $^{12}$CO v=1-0 P(3)-P(9) lines showing the absorption component super imposed on the emission lines. The colored regions show the velocity ranges used for computing integrated line fluxes and creating rotational diagrams (see section 3.5). All the CO lines shown here were re-grid to the same velocity grid to enable an equivalent calculation of the line fluxes between CO lines in the shown velocity regions. The colored regions correspond to velocities of $\pm$14-19 km s$^{-1}$ (red), $\pm$9-14 km s$^{-1}$ (blue), and $\pm$4-9 km s$^{-1}$ (green).}
    \label{fig:line_gal}
\end{figure*}

The spectrally resolved line profiles of the P(3)-P(9) v=1-0 CO lines are shown in Figure \ref{fig:line_gal}. In each of these line profiles, the CO emission is only detected at the $\ge$3$\sigma$ level between velocities of -20 and 20 km s$^{-1}$. The negative velocity side ($<-20$ km s$^{-1}$) of each line may be affected by tellurics as that side of each line was adjacent to  telluric features. To get the highest signal-to-noise line profile for kinematic modeling, we median stacked all of the CO lines shown in Figure \ref{fig:line_gal}. Prior to stacking, we re-grid all of the CO lines shown in Figure \ref{fig:line_gal} to the same velocity axis with their absorption component centered on a velocity of 0 km s$^{-1}$.

To determine the geometric parameters of the emitting CO, we fit a simple Keplerian disk model, similar to the ones in \cite{Pontoppidan08} and \cite{Jensen24} to the median CO v=1-0 line profile from our CRIRES+ observations. This model consists of a vertically flat circumstellar disk with constant surface density that covers a grid of discrete points in radius and azimuth. At each point in the grid, we calculate the observed Keplerian velocity based on the stellocentric distance of the grid point, the azimuthal angle of the grid point, and the disk inclination from 
\begin{equation}
    v=\sqrt{\frac{GM_*}{r}}sin(i)sin(\theta)
\end{equation}

where $G$ is the gravitational constant, $M_*$ is the stellar mass of HD 131488, $r$ is the stellocentric distance of the gas, and $i$ is the disk inclination where 90$^{\circ}$ is edge-on and 0 $^{\circ}$ is face-on, and $\theta$ is the azimuthal angle of the grid point where $\theta=0$ corresponds to the grid point directly along the line of sight to the star. We sampled the grid with 0.1 au grid steps radially from the disk inner radius  ($R_{in}$) to the outer radius ($R_{out}$) and 0.5 degrees azimuthally, spanning 0 to 360 degrees in azimuth. We then added the total number of the grid points at a given velocity to create a model spectrum. By doing so, this model assumes that the CO emission is optically thin. We did not sum each grid point equally, and like in \cite{Pontoppidan08} and \cite{Jensen24}, we introduced a weight on each grid point based on its radial distance from the star such that the intensity of the emission depends on the gas radial distance from the star. This effectively acts as a radial surface brightness gradient in the model. The weight on each grid point was treated as a power-law such that $W\propto r^{a}$, where $W$ is the weight of a grid point, $r$ is the radial distance from the star of each grid point, and $a$ is the power-law index that we leave as a free parameter in our model. Each grid point is also weighted by its geometric area ($R dr d\theta$).

We also included a narrow central absorption component in our kinematic model to capture the effect of the absorption lines super imposed on the emission lines as seen in the CRIRES+ data. We assumed that the absorption line is Gaussian in shape and has a line width set only by the average line-of-sight kinetic temperature of the gas by 
\begin{equation}
    \Delta v=\sqrt{\frac{2kT_{kin}}{m}}
\end{equation}
where $\Delta v$ is the line width in velocity units, $k$ is the Boltzmann constant, $T_{kin}$ is the kinetic temperature of the gas, and $m$ is the mass of a CO molecule. We then gave the absorption component a peak optical depth to scale the depth of the absorption feature. The kinetic temperature of the gas, which sets the line width, and the peak optical depth of the absorption component were left as free parameters. We convolved the absorption line to the spectral resolution of the observations which we determined by fitting sky models of the telluric  lines with the TelFit package \citep{Telfit14} to the CRIRES+ data. This gave a spectral resolution for the observations of R=104,000. We took our CO kinematic model spectrum and multiplied it by $e^{-\tau(v)}$ where $\tau(v)$ is the Gaussian absorption line profile as a function of velocity. The absorption line in our model is centered at a velocity of 0 km s$^{-1}$. The model line profile was also convolved to this resolution to match the data before multiplying by the absorption component. This combination of kinematic model and absorption component was used to produce a model spectrum to compare to the median line profile of the CO lines shown in Figure \ref{fig:line_gal}. 

\subsection{Spectro-astrometic model}
In addition to modeling the line profile of the CO lines detected by CRIRES+, we also simultaneously modeled the SA signal we extracted from our CRIRES+ observations. To generate the SA model, we used the same radial and azimuthal grid of Keplerian velocities of the disk as we used for the line profile model. We then used this grid to compute the SA signal by measuring the distance of each grid point from the star and projecting this distance to the plane of the sky. Since the slit was aligned with the position angle of the disk in our observations, we did not need to add an additional projection of the distance of each grid point along the slit axis. We then computed a weighted sum of all of the grid points within each velocity bin using the power-law profile for the CO emission intensity as a function of stellocentric distance described above as the weight for each point in the grid. The SA signal in each velocity bin was then computed as
\begin{equation}
    \Delta X=\frac{\sum_{i=0}^{N} W_i*d_i }{\sum_{i=0}^{N} W_i}
\end{equation}
where $\Delta X$ is the spectro-astrometric centroid offset, $i$ corresponds to a specific grid point, $N$ is the total number of grid points in each velocity bin, $W_i$ is the weight of each grid point based on the power-law profile, and $d_i$ is the projected distance from the star of each grid point. We then used our simple Keplerian disk model to compute equation 3 for each of the velocity bins in the CRIRES+ data to get a model of the SA signal that covers the same velocities as the extracted SA signal. 

The final step we used to generate a model SA signal was to correct for the line-to-continuum ratio of the CO line because the stellar PSF continuum can act to ``drag" the centroid offset towards zero. The amount that the centroid offset is affected depends on the line-to-continuum ratio and we used the formulism in \cite{Pontoppidan08} to correct for this effect, which is given as 
\begin{equation}
    \Delta x=\Delta X\left(1+\frac{F_c(v)}{F_l(v)}\right)^{-1}
\end{equation}
where $\Delta X$ is the corrected centroid offset, $\Delta x$ is the measured centroid from the Gaussian fitting, $F_c(v)$ is the flux of the stellar continuum as a function of velocity, and $F_l(v)$ is the flux of the CO line as a function of velocity. Our SA model does not have any continuum, so it outputs $\Delta X$, but we need to compute $\Delta x$ to compare directly to the observations. We then used the continuum-to-line ratio as a function of velocity based on the CRIRES+ data and equation 4 to correct the model for the line-to-continuum ratio and compute $\Delta x$. For this correction, we used the stellar continuum flux that we used to produce the continuum subtracted spectrum in Figure \ref{fig:Final_spec} for the $F_c(v)$ term, and we used the model line profile scaled to the flux of the P(3) line as the $F_l(v)$ term. We used the model line profile instead of the data because we find that using the P(3) line profile directly from the CRIRES+ data introduces noise into the SA model, but our line profile model is noiseless. Once we corrected for the line-to-continuum ratio, we then convolved our SA model to the spectral resolution of our CRIRES+ observations to compare it with the data. 

\subsection{Model Fitting Results}
 We jointly fit the median $^{12}$CO v=1-0 line profile along with the SA signal of the $^{12}$CO v=1-0 P(3) line from our CRIRES+ data. The free parameters of our joint line profile and SA model are the inner radius of the gas ($R_{in}$), the outer radius of the gas ($R_{out}$), the radial power-law exponent ($a$) for the intensity of the CO emission as a function of stellocentric radius, the peak optical depth of the absorption line ($\tau$), the kinetic temperature of the absorbing CO ($T_{kin}$), a scaling factor to match the integrated line flux of the median line, and the mass of the central star HD 131488 ($M_*$). We pinned the inclination of the disk in our model to the value determined by modeling the dust continuum observations from ALMA \citep{Zawadski26}, as listed in Table \ref{tab:system_params}. We used the emcee package \citep{Mackey13} to infer the posterior distributions of each free parameter by running a Markov Chain Monte Carlo (MCMC). In the MCMC, we used a Gaussian likelihood function that depends on the $\chi^2$ values from both the line profile fit and the SA fit. We summed together the $\chi^2$ values from both fits in the likelihood function and the two $\chi^2$ values were weighted equally and were computed over the same velocity points. The uncertainties on the spectral and SA data points were determined by computing the standard deviation of the median line profile and SA signal in regions outside of the line and SA signal between velocities of 35-55 km s$^{-1}$. 

\begin{figure*}[htbp!]
    \centering
    \includegraphics[scale=0.7]{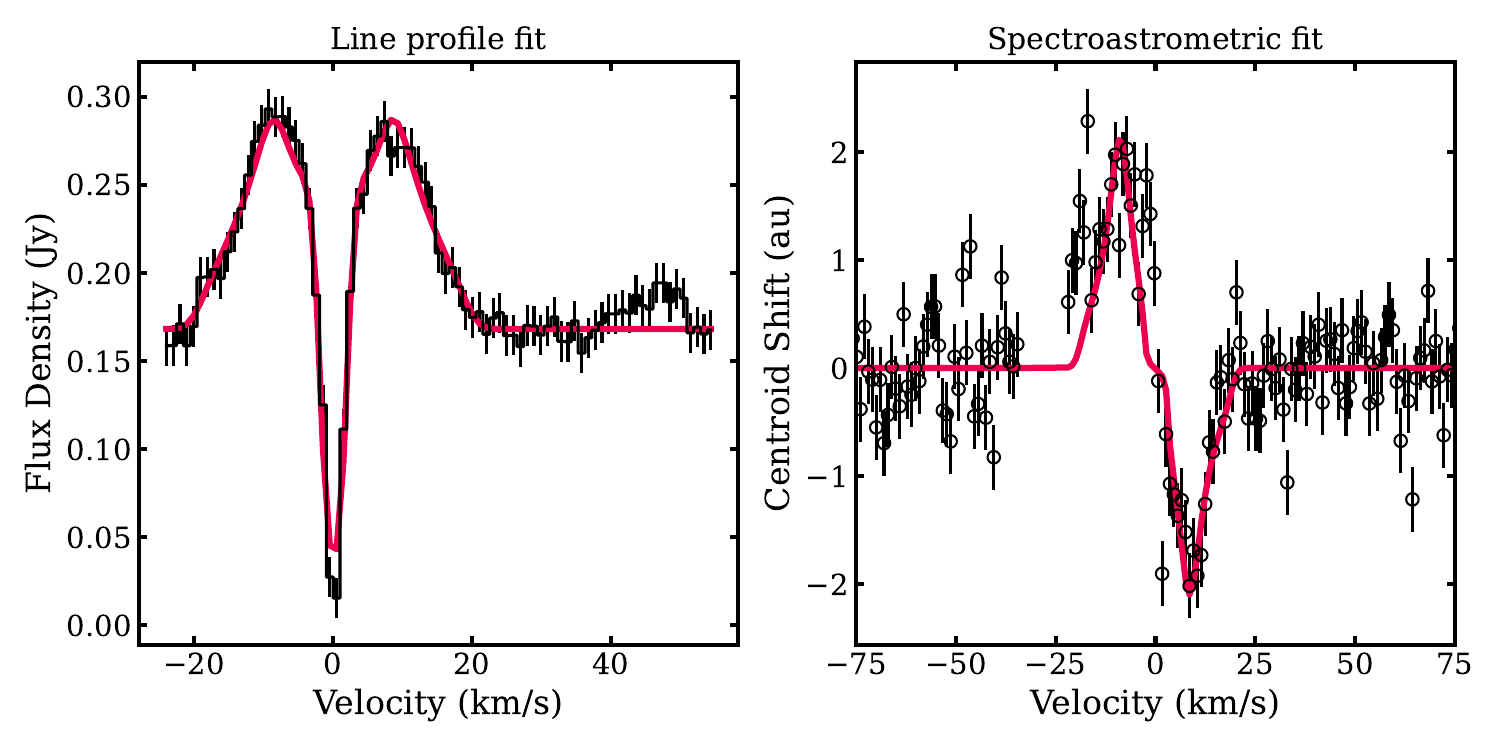}
    \caption{Left: Median line profile of the $^{12}$CO v=1-0 P(3)-P(9) lines (black) with the best fit Keplerian disk model plus absorption component shown in red. Right: Spectro-astrometric profile of the $^{12}$CO P(3) line (black) showing the characteristic SA signature of a circumstellar disk with the best fit SA model (red).}
    \label{fig:line_fit}
\end{figure*}

\begin{figure*}[htbp!]
    \centering
    \includegraphics[scale=0.45]{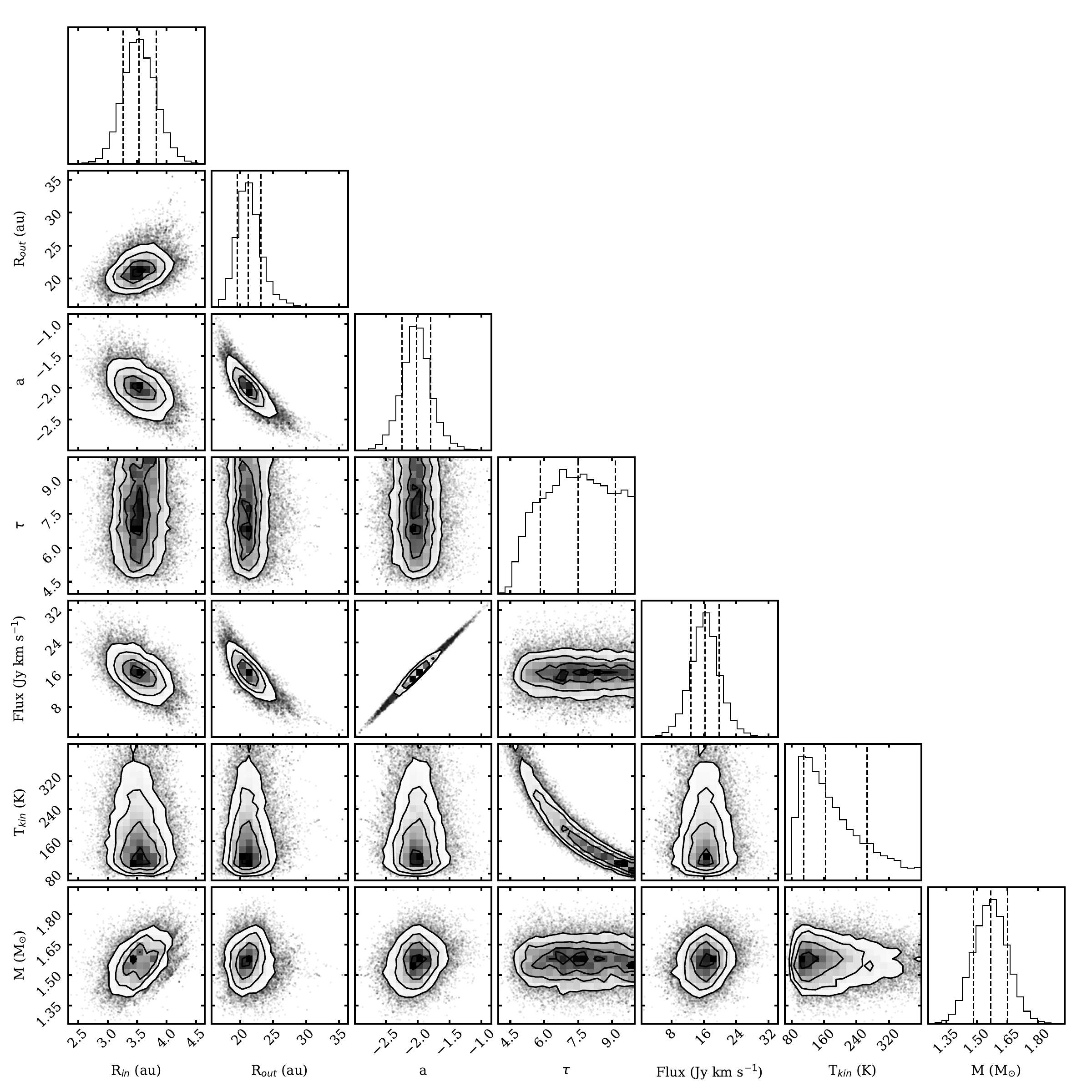}
    \caption{Posterior distributions of each of the free parameters of our CO line profile and SA modeling of the CRIRES+ data of HD 131488. The three vertical lines shown in the posterior distribution of each parameter are the 16th, 50th, and 84th percentiles of distributions.}
    \label{fig:Corner_plot}
\end{figure*}

We used uniform priors on all of the free parameters. We used 90 walkers in the MCMC and ran it for 3000 steps with the first 1500 steps thrown out as the burn-in and we inspected the chain for convergence. The median values of the posterior distribution from the MCMC run and their 16th and 84th percentile uncertainties are shown in Table \ref{tab:line_params}. The best fit line profile model and SA model are shown in Figure \ref{fig:line_fit} and the posterior distributions of each parameter are shown in Figure \ref{fig:Corner_plot}. 

\begin{table}[htbp!]
    \centering
    \caption{Best fit disk parameters}
    \begin{tabular}{c c c}
    \hline
     Parameter & Best Fit (median) & Prior Range  \\
     \hline
         $R_{in}$ (au)& 3.5$^{+0.3}_{-0.3}$ & (0.1, 20) \\ 
         $R_{out}$ (au) &21$^{+2}_{-2}$ & (1, 70) \\
         $a$ &-2.0$^{+0.2}_{-0.2}$ & (-3, 3) \\
         Peak $\tau$&$>$5 & (0, 10) \\
         Line Flux (Jy km s$^{-1}$) &16$^{+4}_{-4}$ & (1, 70) \\
         $T_{kin}$ (K) & 164$^{+102}_{-53}$ & (50, 400) \\
         M$_*$ (M$_{\odot}$) &1.56$^{+0.084}_{-0.085}$ & (0.1, 3) \\
    \hline
    \end{tabular}
         
                \begin{minipage}{7 cm}
    \vspace{0.2cm}
    
     \textbf{Note.} The prior range is shown as  (lower end, upper end) and the ones in parentheses are uniform. The upper limit on the peak $\tau$ is such that 95$\%$ of the values in the posterior distribution are above this value.  
    \end{minipage}

    \label{tab:line_params}
\end{table}

Both the median kinematic line profile and the SA signal are symmetric within the uncertainties, between the red- and blue-shifted sides and both appear to be well explained by the Keplerian circumstellar disk model. Asymmetries and non-Keplerian structures in the SA signal of ro-vibrational CO lines in protoplanetary disks have been detected and attributed to phenomena such as excess CO emission from a circumplanetary disk of an orbiting inner planetary companion \citep{Brittain19} as well as a disk wind that drives gas outward from the central star \citep{Pontoppidan11}. Given lack of asymmetries in the CO line profile and SA signal for the HD 131488 disk, we do not find similar evidence for an inner circumplanetary disk or a disk wind for this system. In the case of the circumplanetary disk in the HD 100546 system, the SA signal was time variable and in some epochs did not exhibit the asymmetric SA profile \citep{Brittain19}. With just one epoch of CRIRES+ data for HD 131488, we cannot  rule out a time variable SA signal. 

\subsubsection{Disk Radial Extent}
Our CRIRES+ observations and kinematic modeling allow us constrain the inner radial extent of the fluoresced warm gas in this system for the first time. High-resolution ALMA observations found that the pure rotational CO emission (J=3-2) at millimeter wavelengths spans from a stellocentric radius of 160 au inward to 6 au \citep{MacManamon26}. Our modeling of the CRIRES+ kinematics of the ro-vibrational CO for this system sets an inner radius of the CO of 3.5$\pm$0.3 au, which is further in than the inner radius of 6 au determined from ALMA observations. We are likely more sensitive to this inner warm CO in our CRIRES+ observations than with ALMA because the transition strengths of the ro-vibrational lines are orders of magnitude stronger than those of the pure rotational lines in the millimeter and the warm excitation temperatures of the gas in inner disks can make the near-IR transitions better probes of warm gas than the millimeter transitions (e.g. \citealt{Pontoppidan02}). Our outer radius of 21 au is smaller than the 160 au CO disk seen with ALMA, but this outer radius of 21 au does not represent the outer extent of the entire gas disk. Instead, the outer radius from our model represents a cutoff of the emitting region that is required for our chosen power-law intensity profile. The choice of the power-law profile is motivated by high-spectral resolution observations of fundamental CO emission in protoplanetary disks (e.g. \citealt{Jensen24}), however, a different parameterization in the radial intensity profile could result in a different outer radius from our model. With the current power-law assumption, a larger $R_{out}$ changes the shape of the line profile and increases the amplitude of the SA signal such that it is inconsistent with the data which is why $R_{out}$ is formally well constrained by our modeling. Although the outer radius of the warm CO from our model is well constrained, it should be interpreted with caution because the assumption we made about the power-law intensity profile makes $R_{out}$ dependent on our model choice.

\subsubsection{Dynamical Stellar Mass}
From our joint fit to the SA signal and stacked CO line profile, we put a constraint on the stellar mass by pinning the disk inclination to the inclination determined from the dust continuum modeling of high resolution ALMA continuum data, $\sim$85$^{\circ}$ \citep{Zawadski26}. Our constraint of M$_*=1.56^{+0.084}_{-0.085}$ M$_{\odot}$ (see Table \ref{tab:line_params}) is $\sim$2-3$\sigma$ lower than the mass determined from modeling photometry with stellar evolution models, which is 1.804$\pm0.016$ M$_{\odot}$ \citep{Marino26}. In performing the modeling of the stacked CO line profile and SA signal, we find that pinning the inclination is what enables the tight constraint on the stellar mass from our fit. When performing the fit without pinning the inclination, the constraint on the stellar mass changes to M$_*=1.7^{+0.4}_{-0.2}$ M$_{\odot}$ because the inclination is not well constrained from our CRIRES+ observations alone. This makes the derived stellar mass consistent with the value determined from photometry. By fitting our line profile and SA signal without pinning the inclination, the only constraint we get on the inclination from our CRIRES+ data is that it is $i>53^{\circ}$. We find that the best fit inner CO radius remains unchanged by allowing the inclination to be a free parameter and the stellar mass is the only parameter affected by our choice of inclination. The inner CO radius determined from our model appears to be robust against the assumption of the disk inclination. Any potential misalignment between the inner CO and the outer dust continuum would, however, result in an incorrect stellar mass reported by our modeling. The stellar mass determined from our kinematic modeling, however, is consistent with the dynamical stellar mass determined from modeling the spatially resolved observations of the CO in the outer disk from ALMA which is found to be 1.66$^{+0.04}_{-0.05}$ M$_{\odot}$ (Grajales et al. in prep.). 

\subsection{Rotational Temperature and Mass of the CO}
With the detection of multiple ro-vibrational CO lines, we can make rotational diagrams based on the observed line fluxes to measure the rotational temperature of the gas. Assuming that the rotational populations of the CO are described by a Boltzmann distribution, the flux of an optically thin emission line can be written as
\begin{equation}
    F=\frac{h \tilde{\nu} c A_{ul} \Omega g_u N_{tot}}{4\pi Q_{rot}(T_{rot})} e^{-E_r/kT_{rot}}
\end{equation}
where $\tilde{\nu}$ is the wavenumber of the transition, $A_{ul}$ is the Einstein A coefficient, $\Omega$ is the emitting area, $g_u$ is the statistical weight of the upper state of the transition (2J+1), N$_{tot}$ is the total column density of CO, $E_r$ is the energy of the upper rotational state of the transition, $Q(T_{rot})$ is the partition function, and T$_{rot}$ is the rotational temperature. We can then make a rotational diagram by plotting  $\ln $$(F/gA_{ul}\tilde{\nu})$ versus the energy of the corresponding upper rotational state of the transition (e.g. \citealt{Goldsmith99}). Note that we plot the upper rotational state energies and not the total upper line energies because the gas is fluorescently excited so the vibrational states are not thermally excited and thus the vibrational and rotational temperatures are decoupled \citep{Brittain07, Worthen25, Lu26}. If the rotational states of CO are populated according to a Boltzmann distribution and the emission is optically thin, then the rotational diagram will produce a straight line where the slope is equivalent to -$1/T_{rot}$ (e.g. \citealt{Brown13}). 

\begin{figure*}[htbp!]
    \centering
    \includegraphics[scale=0.35]{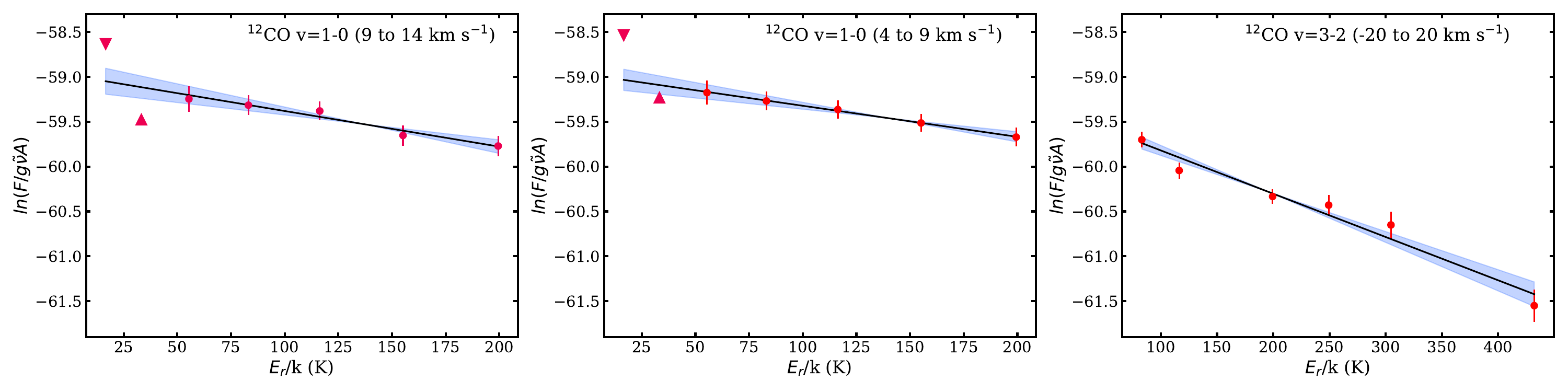}
    \caption{Left: rotational diagram for the $\pm$(9-14) km s$^{-1}$ regions (blue regions in Figure \ref{fig:line_gal}) with best fit line (black).  The blue colored region shows the 1$\sigma$ uncertainty on the best fit line. The two lowest energy lines in the rotation diagram (P(3) and P(4)) were not used to fit the lines and are shown as upper and lower limits respectively. Middle: same as left but for the $\pm$(4-9) km s$^{-1}$ regions (green regions in Figure \ref{fig:line_gal}). Right: Rotational diagram for the $^{12}$CO v=3-2 lines detected our CRIRES+ data integrated over the entire line width from -20 to 20 km s$^{-1}$. The v=3-2 lines shown here are the R(4), R(5), R(7), R(8), R(9), and R(11). Note that all figures share the same y-axis. }
    \label{fig:Rot_dia}
\end{figure*}

The line fluxes for each of the $^{12}$CO v=1-0 emission lines is affected by a narrow absorption component, so the total line flux integrated over the full velocity range of the line would be reduced. Because of this, we compute the line fluxes of each of the $^{12}$CO v=1-0 lines in velocity regions outside of the narrow absorption feature. These velocity regions are shown as the colored areas in Figure \ref{fig:line_gal}. The  colored regions (red, blue, and green) correspond to the same velocity ranges on the red- and blue-shifted side of all of the lines and each region spans a width of 5 km s$^{-1}$ in velocity. The velocities covered by each region are $\pm(14-19)$ km s$^{-1}$ (red), $\pm(9-14)$ km s$^{-1}$ (blue), and $\pm(4-9)$ km s$^{-1}$ (green).
All of the lines were re-grid to the same velocity axis with the peak of the absorption component centered at a velocity of 0 km s$^{-1}$ to ensure that we are integrating the line fluxes over equivalent velocities for each line. We find that the red region is severely impacted by tellurics, as multiple lines have their negative velocity wings in strong telluric features which affects the integrated line flux in that region. For example, The P(9) line has an artifact produced by telluric contamination at $\sim$-20 km s$^{-1}$ as shown in Figure \ref{fig:line_gal}. Because of this, we only present the rotational diagrams of the blue and green regions.

The rotational diagrams for the two velocity regions are shown in Figure \ref{fig:Rot_dia}. We do not use the P(3) or P(4) in fitting the lines in the rotation diagrams to determine the rotational temperature. This is because the P(3) line overlaps in wavelength with the $^{12}$CO v=2-1 R(3) line as shown in Figure \ref{fig:Final_spec} and thus the flux from that line would be over estimated compared to the others, so its integrated line flux is shown as an upper limit in the rotational diagrams in Figure \ref{fig:Rot_dia}. The P(4) line is heavily affected by tellurics, as shown in Figure \ref{fig:Tell_spec}, and when comparing its line flux to the other lines, we find that the total P(4) line flux is about half of what is seen in the JWST/NIRSpec data \citep{Lu26}. We therefore do not include it in the fitting of the rotational diagram and treat it as a lower limit on the line flux in Figure \ref{fig:Rot_dia}. 

Only the five points shown as red dots in Figure \ref{fig:Rot_dia} are used in the line fitting of the rotational diagrams. The uncertainties on each point in the rotation diagram are calculated by first measuring the standard deviation of a line-free spectral region beyond 35 km s$^{-1}$ of the line center of each v=1-0 line. We then treat that uncertainty as the uncertainty on each spectral point in the CO line and propagate it through the calculation of integrating the line flux over velocity and computing the y-axis quantity ($ln(F/g\tilde{v}A)$) of the rotational diagrams. We also add the absolute flux calibration uncertainty from JWST/NIRSpec (3$\%$) in quadrature \footnote{\url{https://jwst-docs.stsci.edu/jwst-calibration-status/nirspec-calibration-status/nirspec-fixed-slit-calibration-status}}, since we used the NIRSpec spectrum to flux calibrate our CRIRES+ data, to the uncertainty on each spectral point before propagating it through the line flux calculation.

We also integrate the line fluxes of the $^{12}$CO v=3-2 lines to create a rotational diagram to compare with the temperature derived from the $^{12}$CO v=1-0 lines. We use all of the $^{12}$CO v=3-2 that are in our data, which are the R(4), R(5), R(7), R(8), R(9), and R(11) lines. Because the v=3-2 lines do not have the central absorption component, we integrate the line fluxes across the entire velocity range of -20 to 20 km/s. The rotational diagram is shown in the right panel of Figure \ref{fig:Rot_dia}.

From the best fit lines to each of the rotational diagrams, we find that the rotational temperatures of the CO for the two velocity regions (9-14 and 4-9 km s$^{-1}$) are 258$^{+90}_{-54}$ K and 273$^{+96}_{-61}$ K respectively. The rotational temperature we derive from the $^{12}$CO v=3-2 lines is 207$^{+28}_{-23}$ K. The rotational temperature of the CO derived from the v=1-0 lines and the v=3-2 lines is consistent within the uncertainties. To estimate a lower limit on the total mass of CO between 3.5 and 21 au, we first sum the CO masses that we calculate within each rotational population of the v=1 state, excluding the lines that were not used in the rotational diagram fitting (P(3) and P(4)). This summation gives a conservative lower limit to the total CO mass distributed across the rotational levels of the v=1 state and is done instead of using the rotational temperatures derived from the rotational diagrams since UV fluorescence may alter the rotational populations of CO \citep{Matra18}. If the rotational levels of the ground vibrational state are thermalized, however, then their rotational states will be populated by Boltzmann distributions, even in the case of UV fluorescence \citep{Matra18}. Because of quantum selection rules that $\Delta J=\pm1$, the rotational temperature of the CO still provides an estimate of the rotational temperature of the UV fluoresced CO \citep{Brittain07, Brown13} as long as the rotational states of the ground vibrational state are thermalized. If the ground vibrational state rotational populations are not thermalized, then the rotational temperature reported here will not be representative of the gas temperature. Our rotational diagrams for the transitions we detect are well explained by a Boltzmann distribution, however, we only detect up to the R(11) line in emission, so it is possible that higher J states that we do not probe are not thermalized and the rotational temperature could be lower than what we report.

We also accounted for the fraction of CO molecules in the v=1 state in the calculation of CO mass, as the gas is fluorescently excited \citep{Lu26} and the fraction of CO molecules in the v=1 state depends on the distance of the CO from the star and the stellar UV field \citep{Brittain07}. We used the fluorescent excitation model from \cite{Worthen25} and the stellar parameters of HD 131488 from \cite{Brennan24} to calculate the fraction of CO molecules in the v=1 state at a distance of 3.5 au. The fluorescent excitation model balances the excitation and de-excitation processes from UV fluorescence via the A-X band UV transitions of CO given a input stellar spectrum and the distance of the CO from the star. For the stellar flux, we used an ATLAS 9 photosphere model \citep{Kurucz03} with the stellar parameters of HD 131488 from \cite{Brennan24}. See \cite{Worthen25} for a complete description of the model. We find that the fraction of CO molecules in the v=1 state at 3.5 au from HD 131488 is 6.5$\times10^{-3}$. 

We then use this fraction of CO molecules in the v=1 state together with the summed CO masses from both velocity ranges to estimate a lower limit on the total CO mass, which we find to be 3.6$\times10^{-5}$ M$_{\oplus}$. This CO mass should be thought of as a lower limit to the total CO mass between 3.5 and 21 au around HD 131488 as we are only sensitive to the CO that is fluoresced by UV photons, which may only be the upper region of the gas disk \citep{Brittain07, Worthen25}. We also assumed the CO emission was optically thin making this mass a minimum CO mass. Further, this mass is only computed from the wings of the CO lines and does not include the absorbed out central line flux and we have also made the simplification of only computing the fraction of CO in the v=1 state at the gas inner radius thus underestimating the total CO mass. Even though the mass we compute is a minimum CO mass between 3.5 and 21 au, it still serves as a useful comparison to secondary gas evolution  models (see Section 4).

\section{Discussion}

\subsection{What is the origin of the inner CO?}
Our modeling of the CRIRES+ kinematics of the ro-vibrational CO emission from the HD 131488 debris disk indicates that the gas extends inwards to a stellocentric distance 3.5$\pm$0.3 au. This result suggests that the CO does not extend inwards to sub-au stellocentric distances like the CO in some protoplanetary disks \citep{Dickson-Vandervelde25}. Using a bolometric luminosity of 11.8 $L_{\odot}$ for HD 131488 \citep{Matra25}, this distance of inner radius of CO of 3.5 au from HD 131488 corresponds to an equivalent distance, based on radiative equilibrium temperature, of $\sim$1 au in the Solar System. This CO is therefore present in a region around HD 131488 analogous to the Solar System's terrestrial planet zone, based solely on radiative equilibrium temperature. Here, we explore possible scenarios that could give rise to this terrestrial planet zone gas, including viscous evolution of second-generation gas released from minor bodies in the outer planetesimal belt of the system ($\sim$100 au) as well as the possibility that the gas is of a primordial origin from the parent protoplanetary disk. 

Models of second-generation gas production in debris disks predict that, if CO self-shielding and shielding  by atomic carbon against photodissociation are sufficiently high, then second-generation CO can viscously spread inwards towards the central star \citep{Kral19, Marino20}. Depending on the strength of the viscosity, the CO and atomic carbon gas can spread inwards until it reaches the central star, forming an accretion disk of second-generation gas \citep{Kral19, Marino20}. As this secondary gas spreads inwards, it may cross the orbits of existing planets, where it could accrete onto a planet either changing its atmospheric composition or seeding it with a secondary atmosphere \citep{Kral20}. Rocky planets of a wide mass range (Mars mass up through 5$M_{\oplus}$) in these secondary gas disks have been shown to efficiently accrete gas and, if massive enough, they can stop gas from reaching the central star, creating cavities in the gas disk \citep{Kral20}. 

To test whether the warm terrestrial zone gas detected with CRIRES+ can be explained by viscously evolving gas released from the outer planetesimal belt, we use the EXOGAS code \citep{Marino20, Marino22} to model the radial evolution of the CO in the HD 131488 disk under the assumption that the gas is of second-generation. EXOGAS simulates the radial evolution of the surface density of CO and atomic carbon, a photodissociation product of CO, based on a constant CO production rate, the gas viscosity (parameterized by $\alpha$), the stellar luminosity, and the location and width of the planetesimal belt producing the gas. For the location and width of the planetesimal belt ($R_{belt}$ and $\Delta R$), we use the constraints on the spatial distribution of the millimeter sized dust grains probed by high-angular resolution ALMA observations \citep{Han26}, which give $R_{belt}=93$ au and $\Delta R=43$ au. We also set the ionization fraction of atomic carbon to be 0.1, which is a value expected for shielded disks with high carbon densities \citep{Marino20}. With these parameters, we run a grid of EXOGAS models where we vary the CO production rate ($\dot{M}_{co}$) and $\alpha$, which sets the viscosity as $v=\alpha c_s^2/\Omega_k$ where $v$ is the viscosity, $c_s$ is the gas sound speed, and $\Omega_K$ is the Keplerian orbital speed. Our grid spans values of the exponent for $\alpha$ ($\alpha=10^p$) from $p=$-4 to -1 in steps of 0.15 and the exponent for $\dot{M}_{co}$ (in the form of 10$^{q}$ $M_{\oplus}$/Myr) from values of $q=$-3 to -1  in steps of 0.15. 

\begin{figure*}[htbp!]
    \centering
    \includegraphics[scale=0.5]{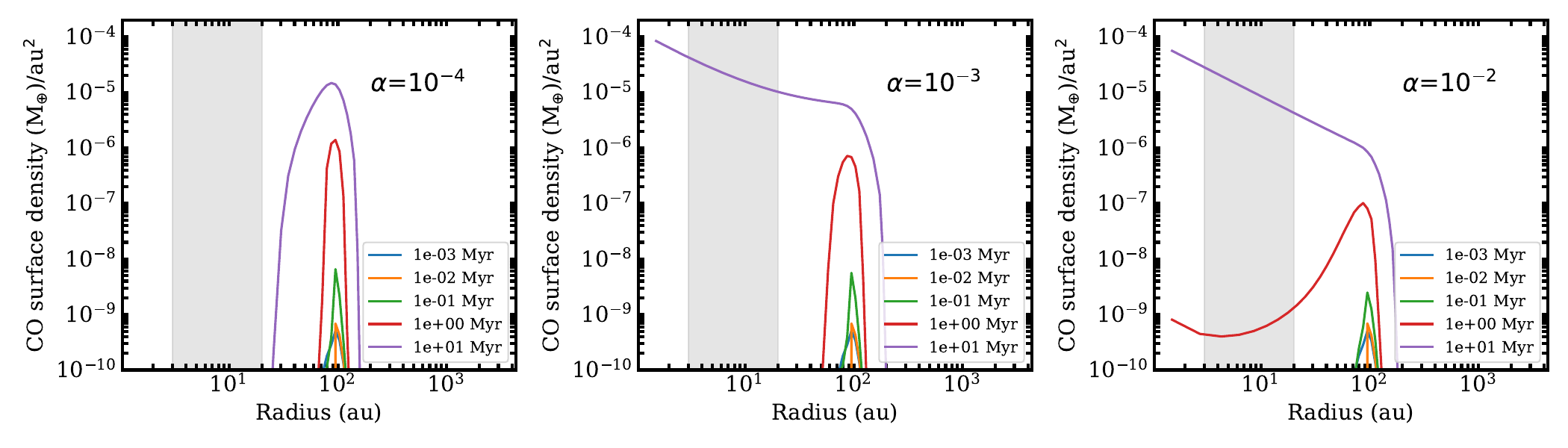}
    \caption{Left: EXOGAS model radial surface densities of CO in the HD 131488 system for a CO production rate of 5$\times10^{-2}$ ${M}_{\oplus}$/Myr and  $\alpha=1\times10^{-4}$. Each colored line shows the surface density at a given time. The gray colored region shows the stellocentric distances of the fluoresced CO from our analysis of the CRIRES+ line profiles and SA signal.  Middle: same as left but for  $\alpha=1\times10^{-3}$. Right: same as left but for $\alpha=1\times10^{-2}$. }
    \label{fig:Surface_den}
\end{figure*}

EXOGAS only includes photodissociation from the interstellar radiation field (ISRF) and does not include photodissociation from the central star. Because the CO we detect with CRIRES+ extends inwards to 3.5 au, the stellar photodissociation will be stronger than that of the ISRF at these close distances \citep{Brennan24} and will thus affect the evolution of the gas close to the star. We therefore include stellar photodissociation in the EXOGAS model to test if it can reproduce the lack of CO emission from stellocentric distances inward of 3 au. We follow the procedure of \citep{Brennan24} to add the stellar photodissociation rates into the EXOGAS model. We calculate the photodissociation rate of unshielded CO as 
\begin{equation}
    k=\int\sigma(\lambda)I(\lambda)d\lambda
\end{equation}
where $\sigma$ is the photodissociation cross section of CO taken from \cite{Visser09}, $I(\lambda)$ is the stellar flux in photon based intensity for which we use a ATLAS9 stellar photosphere model \citep{Kurucz03} with the stellar parameters of HD 131488 determined by \cite{Brennan24}. We then account for self-shielding of CO and shielding by atomic carbon by multiplying the photon based intensity of the central star inside the integral of equation 6 by $\exp$$(-\sigma_{CO}N_{CO})$ and $\exp$$(-\sigma_{C}N_{C})$ where $N_{CO}$ and $N_C$ are the radial column densities of CO and atomic carbon and $\sigma_C$ is the photoionization cross-section of atomic carbon as a function of wavelength from \citep{Vandishoek88}. The radial column densities of CO and atomic carbon along the line of sight to the central star are computed at each radial point in the model using the CO and C surface densities at each timestep in the model through the relation from \cite{Brennan24} of $N_i\propto\Sigma_i\frac{\Delta R}{2H}$ where $\Sigma_i$ is the surface density of the respective species $i$, $\Delta R$ is the width of each ring (i.e. distance between each radial grid point), and $H$ is the disk scale height. Like in \citep{Brennan24}, we assume that the scale height is defined with a constant aspect ratio $h=H/r$ and $h=0.1$. After incorporating stellar photodissociation into the EXOGAS model, we run the grid of simulations with varying $\alpha$ and $\dot{M}_{co}$ to the 16 Myr age of the system \citep{Pecaut16}.

The evolution of the radial surface density of the CO over time from a subset of our EXOGAS model grid is shown in Figure \ref{fig:Surface_den}. For the  model results shown in Figure \ref{fig:Surface_den}, the CO input rate is the same (5$\times10^{-2}$ ${M}_{\oplus}$/Myr) for each of the 3 panels, but they have different input viscosity parameters $\alpha$. This shows that for this CO input rate and $\alpha=10^{-4}$, the CO does not spread viscously inward enough from its release point to have a CO surface density above 10$^{-10}$ $M_{\oplus}$/au in the region where we determined the CO emission detected by CRIRES+ arises (gray shaded region in Figure \ref{fig:Surface_den}). Further, Figure \ref{fig:Surface_den} shows that in the higher-$\alpha$ cases (10$^{-3}$ and 10$^{-2}$), the CO does spread inwards into the region where the warm gas is detected and also spreads all the way inwards to the central star even with the stellar photodissociation incorporated into the model. Stellar photodissociation does not appear to create a lack of CO at distances less than 3 au in our model runs due to strong self shielding and shielding from atomic carbon in the model. 

\begin{figure}
    \centering
    \includegraphics[scale=0.65]{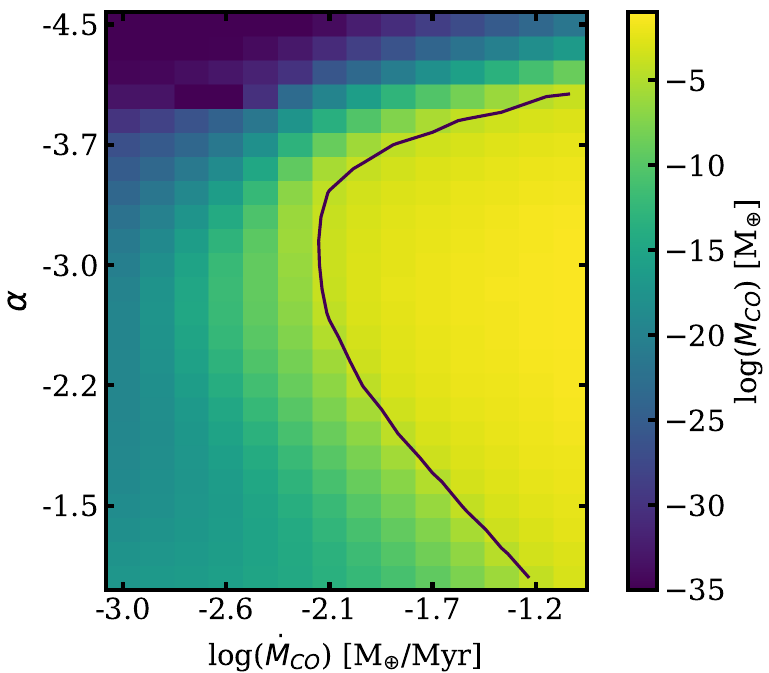}
    \caption{CO mass between 3 and 20 au at the end of each EXOGAS model for the ranges of $\alpha$ and CO production rate ($\dot{M}_{CO}$) values in our model grid. The black solid line shows the minimum mass of CO in the 3 to 20 au range of HD 131488 as determined from our CRIRES+ observations. }
    \label{fig:Model_grid}
\end{figure}

 We compare the CO mass at the end time of each EXOGAS model between stellocentric distances of 3 and 20 au with the minimum CO mass calculated using the CO line fluxes from the CRIRES+ observations to determine if the viscous spreading of secondary gas can explain the minimum warm CO mass determined from our CRIRES+ observations. Figure \ref{fig:Model_grid} shows the CO mass between 3 and 20 au at the end of each EXOGAS model in our grid compared with the minimum warm CO mass determined from the CRIRES+ observations which is overlaid as a black line.
All the models corresponding to grid points that are contained between the black line and the rightmost edge of Figure \ref{fig:Model_grid} are able to reproduce the observed minimum CO mass in the 3 to 20 au range determined from our CRIRES+ observations. For CO production rates up to the maximum one in our model grid (0.1 M$_{\oplus}$/Myr), the EXOGAS results suggest that an $\alpha$ value greater than $10^{-4}$ is required for the gas to viscously spread inwards to provide sufficient CO mass to explain the CRIRES+ observations. This value of $\alpha$ required to explain the inner gas of HD 131488 is generally consistent with population studies of gas evolution in debris disks that require $\alpha$ to be high, between $10^{-3}$ and 0.1 \citep{Kral19, Marino20}, with \cite{Marino20} favoring a value of $\alpha\sim$0.1 to explain the population of gas rich debris disks.   

This modeling demonstrates that the CO observed in the inner regions of the disk could be the result of viscous spreading of gas released from the outer planetesimal belt in a secondary gas origin scenario. An important caveat to this result, however, is that the EXOGAS model predicts a higher atomic carbon-to-CO column density ratio than what is observed with HST UV absorption line spectroscopy of HD 131488 \citep{Brennan24}. \cite{Brennan24} suggest that one possible explanation for the discrepancy between the observed carbon-to-CO ratio and those required by the secondary gas models to enable viscous spreading is that the atomic carbon might be concentrated in a thin vertical layer on top of the CO, which could possible if vertical diffusion is low \citep{Marino22}. In this case, the observed carbon-to-CO ratio from absorption line spectroscopy would serve as a lower limit on the C/CO ratio in the HD 131488 disk. If the carbon is instead well mixed with the CO and not vertically layered, then the HST-derived C/CO ratio would suggest that there is not enough atomic carbon to shield CO against photodissociation to enable this viscous spreading of the outer gas inwards as the EXOGAS model predicts. This would then require additional shielding sources in the disk to enable the viscous spreading, or that the gas is instead produced in situ from parent bodies at  distances of $\sim$3 au. This additional shielding source could be because the gas is of a primordial origin rather than a secondary one, where H$_2$ would shield the CO against photodissociation. 

In addition to the warm gas around HD 131488, warm dust has also been detected in this system. The warm dust around HD 131488 has been interpreted to suggest that there are ongoing collisions between rocky bodies in the inner regions of the system \citep{Melis13}. The blackbody radiative equilibrium temperature for the  warm dust puts it at a stellocentric distance of 0.6 au \citep{Melis13}, which is interior to our CO inner radius. The blackbody distance of the dust based on its temperature has been known to underestimate the size of dust disks (e.g. \citealt{Booth13}). This can be due to small grains being at higher equilibrium temperatures than larger grains since small grains  are inefficient radiators at large wavelengths \citep{Jura98}. We can then estimate the distance of dust in the small grain limit (2$\pi$a $<<\lambda$, where $a$ is the grain radius) using equation 2 in \cite{Worthen24b} for the radiative equilibrium temperature of dust in the small grain limit, the 750 K dust temperature \citep{Melis13}, and the stellar parameters of HD 131488 from \citep{Brennan24}. Doing so, we find that 750 K dust in the small grain limit would be at a stellocentric distance of 1.6 au, which is still interior to our inner radius of CO of 3.5$\pm$0.3 au. Given that the warm dust and inner CO may not be co-spatial, it is not obvious that the two are related. Further, high-resolution ALMA observations of the CO reveal that the gas distribution is continuous from the outer region of the disk (160 au) to inner-most resolution element of the data \citep{MacManamon26}, potentially suggesting that the inner gas seen here with CRIRES+ is connected to the outer gas seen by ALMA either through either viscous evolution of secondary gas or it may be a continuous population of primordial gas.

\subsection{Can an accreting planet create the lack of CO inwards of 3 au?}

Even with photodissociation from stellar photons, we cannot reproduce the lack of CO at stellocentric distances smaller than 3 au, as evidenced by our CRIRES+ observations, with the EXOGAS models. In the cases where the viscosity is high enough for gas to spread from the outer regions to the inner region where we detect the CO with CRIRES+, the CO continues to spread all the way inwards to the central star as shown in Figure \ref{fig:Surface_den}. Similar results were found for a population synthesis of the radial evolution of gas in debris disks, where the atomic carbon and/or CO spread all the way to the central star, creating an accretion disk \citep{Marino20}. Lack of viscous spreading of gas into the central star has also been seen with ALMA observations of atomic carbon in the HD 32297 debris disk which find that the atomic carbon is concentrated in a ring \citep{Cataldi20}. One potential mechanism that could stop the gas from spreading to the central star and forming an accretion disk is that there could be planets between the gas release point and the central star that accrete the second-generation gas as it spreads inwards \citep{ Marino20, Kral20}. 

Planets have been suggested to create gaps in gas disks and decrease the inflow of gas to the central star as the planet accretes the gas as the gas spreads inwards \citep{Lubow06,Kral20}. The ability for a planet to accrete gas from the circumstellar disk and create a cavity ultimately depends on the planet's mass. More specifically, a requirement for a planet to slow the radial flow of gas inwards and create a gap near its orbit is that the planet's Hill radius must be larger than the gas disk's scale height (e.g. \citealt{Lin93, Kuwahara19}). Using this requirement, we can then find the minimum mass of a planet required to create a cavity in the CO distribution at the inner radius of the CO probed by our CRIRES+ observations (3.5 au). Assuming that the gas is in hydrostatic equilibrium, we can write the gas scale height as
\begin{equation}
    H=\left(\frac{kTR^3}{\mu m_H G M_*}\right)^{1/2}
\end{equation}
where $k$ is the Boltzmann constant, $T$ is the gas temperature, $R$ is the stellocentric distance of the gas, $\mu$ is the mean molecular weight of the gas, $m_h$ is the mass of the hydrogen atom, $G$ is the gravitational constant, and $M_*$ is the stellar mass. We then use equation 7 along with the requirement that a planet's Hill radius be larger than the gas scale height to determine the minimum mass planet that could create a gap at 3.5 au, which is 
\begin{equation}
    M_p> 3\left( \frac{kTR}{\mu m_H G M_*^{1/3}}\right)^{3/2} 
\end{equation}
where $M_p$ is the mass of a planet that can create a gap in the disk by accreting the circumstellar gas. The remaining symbols represent the same quantities as in equation 7. Assuming that the gas temperature is equal to the rotational temperature calculated from the CRIRES+ observations (273 K), the mean molecular weight of the gas is $\mu$=14 which corresponds to gas dominated by atomic carbon and oxygen \citep{Matra17}, and using a stellar mass of 1.8 $M_{\odot}$ \citep{Marino26}, we find that the minimum mass planet required to accrete gas and create a cavity at 3.5 au is 10 M$_{\oplus}$. If we assume that the mean molecular weight is instead dominated by CO ($\mu=28$), then the minimum mass planet required to create the cavity at 3.5 au is 4 M$_{\oplus}$. This estimate is compareable to that from \cite{MacManamon26} which estimated the planet mass required to carve a gap in the gas disk based on the ALMA observations of CO which suggested that R$_{in}<$ 6 au. Their estimate in the $\mu$=28 case gave a minimum planet mass of 7 M$_{\oplus}$. Such planets are currently below the detection limits in H-band VLT/SPHERE images of the system \citep{Pawellek24}. The non-detection of planets via high-contrast imaging techniques for this system still allows this scenario of planetary mass objects being the reason that the gas does not spread all the way into the central star. 

\subsection{What about a primordial origin for the gas?}
So far, our modeling of the gas radial profile has assumed that the gas is of secondary origin, where it is released from icy bodies in an outer planetesimal belt and then viscously spreads inwards to the regions where we detect the CO emission from warm fluoresced gas with the CRIRES+. Another possibility is that the gas is not of secondary origin, but is instead of primordial origin where the CO is remnant of the parent protoplanetary disk. Such a scenario has been proposed for the high-CO mass debris disks around A-type stars where the weak EUV and X-ray emission from A-type stars, due to their lack of surface convective zones, can allow for extended gas disk lifetimes beyond 10 Myr if the disk lacks small dust grains to slow gas photoevaporation \citep{Nakatani23}. HD 131488 is one such A-type star system with one of the highest CO masses in any debris disk \citep{Moor17}. The primordial gas disk evolution models that lack small dust grains predict that the gas will first become depleted in the inner regions of the disk ($\lesssim 10$ au), potentially giving rise to gas cavities close to the star \citep{Ooyama25}, which would be qualitatively consistent with our CRIRES+ kinematic analysis of the CO in this system. This would be similar to what is expected in an inside-out disk clearing scenario (e.g. \citealt{Ercalano17, Coleman22}) where a gap opens up in the inner regions of the disk due to stellar photoevaporation.  

A way to distinguish between the primordial and secondary gas scenarios is to determine the H$_2$ abundance of the gas disk. Previous observations with CRIRES+ in K-band searched for H$_2$ absorption along the line of sight to HD 131488 and resulted in a non-detection of H$_2$ and a limit on the CO/H$_2$ ratio of CO/H$_2>$3.09$\times10^{-5}$ \citep{Smith25}. This limit cannot completely rule out an CO/H$_2$ ratio similar to that of the ISM or that is found in protoplanetary disks (e.g. \citealt{France14}). \cite{Smith25}, therefore, could not completely rule out a primordial origin of the gas around HD 131488. 

In the case of the CO around $\beta$ Pictoris, it was found that the CO lines both at millimeter and near-IR wavelengths showed evidence for a sub-thermal excitation of the gas \citep{Troutman11,Matra17}. This sub-thermal excitation was used to suggest that the disk has an elevated CO/H$_2$ ratio compared to the ISM and protoplanetary disks, as H$_2$ can act as a collisional partner with CO which directly affects the CO excitation \citep{Matra17}. For HD 131488, the ro-vibrational CO emission detected by CRIRES+ does not show evidence for sub-thermal excitation as the CO rotational level populations measured from the CRIRES+ data do not show any significant departures from what is expected from a Boltzmann distribution (see Figure \ref{fig:Rot_dia}). This, however, does not rule out a sub-thermal excitation or a secondary gas origin scenario given the uncertainties on the line fluxes, the assumption of optically thin emission, the possibility of other collisional partners such as free electrons to excite the CO, and that fluorescence can act to increase the populations of higher J levels of CO \citep{Matra18} which would directly affect the measured rotational populations in our rotational diagrams. We cannot rule out a primordial or secondary origin for the gas in this disk with our analysis alone. The spatial distribution of the gas as probed by our CRIRES+ observations could be explained by both primordial gas that is being cleared from the inside-out or second-generation gas that viscously spreads inwards towards the central star after it is released from the outer planetesimal belt. 

\subsection{Explanations for the Dynamical Stellar Mass}

As discussed above, our dynamical stellar mass is lower than what is predicted from stellar evolution models \citep{Marino26}, and the choice of the inner disk inclination may be the cause of the discrepancy. Another potential explanation for the discrepancy between the dynamical mass from our CRIRES+ observations and the stellar mass determined from stellar evolution models is that the CO seen by CRIRES+ is subject to radiation pressure, making the gas ``feel" a less massive star. Molecules subject to radiation pressure feel a reduced gravitational force for their given distance based on the ratio of radiation pressure to gravity ($\beta$) \citep{Krivov06} where this reduced radial force is given as 
\begin{equation}
    F=\frac{GM_*(1-\beta)}{r^2}.
\end{equation}

For circumstellar gas around early type stars like HD 131488, the ratio of radiation pressure to gravity ($\beta$) depends on the stellar luminosity and the number and strength of transitions between $\sim$0.1 and 5 $\mu$m \citep{Fernandez06}. Given that this CO around HD 131488 seen by CRIRES+ is fluoresced by stellar UV photons, it must experience a non-zero radiation pressure. The exact value of $\beta$ for the CO seen here would require detailed modeling as the $\beta$ value would depend not only on the line parameters and stellar luminosity, but also the density of the gas in the inner disk as self shielding and collisions between gas molecules could potentially affect the outward radiation pressure \citep{Marino20}. It would require similar modeling to also test if the inner CO cavity inferred from our CRIRES+ observations could be created by radiation pressure from stellar photons. Computing the dynamics to determine if the radiation force can explain the stellar mass discrepancy would require detailed modeling and is beyond the scope of this paper. However, to explain the difference in the dynamical stellar mass from our modeling and the mass from the stellar evolution models, we require a $\beta=0.14$ based on equation 9.

\section{Conclusions}
We present CRIRES+ M4368 observations of fundamental ro-vibrational CO emission from the HD 131488 debris disk and our main findings are:

\begin{itemize}
    \item We detect the $^{12}$CO v=1-0 P(3)-P(9) fundamental  lines in emission and spectrally resolve their line profiles with CRIRES+.

    \item The line profiles and the spectro-astrometric signal from the CO emission are consistent with a circumstellar disk in Keplerian rotation with no evidence for asymmetries at the current SNR of the data.
    
    \item We model the kinematic profile and the spectro-astrometric signal of the CO emission and find that the inner radius of the gas is 3.5 $\pm$0.3 au, which is a distance equivalent in equilibrium temperature to the Solar System's terrestrial planet zone. \\

    \item Through secondary gas radial evolution modeling, we find that the inner gas detected by CRIRES+ could be the result of viscous spreading of CO produced in the outer planetesimal belt if the viscosity parameter $\alpha> 10^{-4}$ and the gas is secondary. 

     \item The kinematic CO profile indicates that the CO does not extend all the way into the central star, as predicted by the radial evolution models and we find that undetected planets of mass greater than 10 M$_{\oplus}$ could accrete the CO before it spreads to the central star.

     \item The inner cavity of the CO could also be explained by the clearing from the inside out of primordial gas, as our data does not show evidence for sub-thermal excitation of the warm CO.

\end{itemize}

KW and CC acknowledge the support from the NASA FINESST program. This work is supported by the National Aeronautics
and Space Administration under Grant No. 80NSSC22K1752
issued through the Mission Directorate. Kadin Worthen’s research was supported by an appointment to the NASA Postdoctoral Program at the NASA Goddard Space Flight Center, administered by Oak Ridge Associated Universities under contract with NASA. A.M.H. gratefully acknowledges support from the National Science Foundation under Grant No. AST-2307920. L.M. acknowledges funding by the European Union through the E-BEANS ERC project (grant No. 100117693). Views and opinions expressed are however those of the author(s) only and do not necessarily reflect those of the European Union or the European Research Council Executive Agency. Neither the European Union nor the granting authority can be held responsible for them.

\software{
This research has made use of the following software projects:
    \href{https://astropy.org/}{Astropy} \citep{Astropy13,astropy18,Astropy22},
    \href{https://matplotlib.org/}{Matplotlib} \citep{matplotlib07},
    \href{http://www.numpy.org/}{NumPy} and \href{https://scipy.org/}{SciPy} \citep{numpy07}
    and
    the NASA's Astrophysics Data System.
}


\bibliographystyle{aasjournal}
\bibliography{mybib}






\end{document}